\documentclass[11pt,a4paper]{article}
\pdfoutput=1
\usepackage{jcappub}

\ifx\pdfoutput\undefined
\usepackage[dvips,bookmarks=false]{hyperref}	
\else
\usepackage{hyperref}	
\fi
\hypersetup{colorlinks,bookmarksopen,bookmarksnumbered,citecolor=blue,
linkcolor=black,pdfstartview=FitH,urlcolor=blue}

\usepackage{graphicx}
\usepackage{amsmath}
\usepackage{amssymb}

\newcommand{\be}{\begin{equation}}
\newcommand{\ee}{\end{equation}}
\newcommand{\beq}{\begin{equation}}
\newcommand{\eeq}{\end{equation}}

\newcommand{\vect}[1]{\boldsymbol{\rm #1}}

\DeclareMathOperator\erf{erf}

\newcommand{\vv}{\textbf{v}}

\makeatletter
\renewcommand{\fnum@table}{\textbf{\tablename~\thetable}}
\renewcommand{\fnum@figure}{\textbf{\figurename~\thefigure}}
\makeatother

\title{What is the probability that direct detection experiments have observed Dark Matter?}

\def\mpi{Max-Planck-Institut f{\"u}r Kernphysik,\\ 
Saupfercheckweg 1, 69117 Heidelberg, Germany}
\def\grappa{GRAPPA, University of Amsterdam,\\ 
Science Park 904, 1098 XH Amsterdam, Netherlands}
\def\sthlm{Oskar Klein Centre for Cosmoparticle Physics, 
Department of Physics,\\ 
Stockholm University, SE-10691 Stockholm, Sweden}

\author[a, b]{Nassim Bozorgnia}
\author[c]{and Thomas Schwetz}

\affiliation[a]{\mpi}
\affiliation[b]{\grappa}
\affiliation[c]{\sthlm}

\emailAdd{n.bozorgnia@uva.nl}
\emailAdd{schwetz@fysik.su.se}

\abstract{In Dark Matter direct detection we are facing the situation
  of some experiments reporting positive signals which are in conflict
  with limits from other experiments. Such conclusions are subject to
  large uncertainties introduced by the poorly known local Dark Matter
  distribution.  We present a method to calculate an upper bound on
  the joint probability of obtaining the outcome of two potentially
  conflicting experiments under the assumption that the Dark Matter
  hypothesis is correct, but completely independent of assumptions
  about the Dark Matter distribution. In this way we can quantify the
  compatibility of two experiments in an astrophysics independent
  way. We illustrate our method by testing the compatibility of the
  hints reported by DAMA and CDMS-Si with the limits from the LUX and
  SuperCDMS experiments. The method does not require Monte Carlo
  simulations but is mostly based on using Poisson statistics. In
  order to deal with signals of few events we introduce the so-called
  ``signal length'' to take into account energy information. The signal length method provides a simple
  way to calculate the probability to obtain a given experimental
  outcome under a specified Dark Matter and background hypothesis.
}

\arxivnumber{1410.6160}

\begin{document}
\maketitle

\section{Introduction}
\label{sec:introduction}

Dark Matter (DM) direct detection experiments search for a tiny
nuclear recoil signal induced by the scattering of a DM particle from
the galactic halo with a nucleus in the detector
\cite{Goodman:1984dc}. In recent years a number of experiments have
reported signals which may be interpreted in terms of DM
\cite{Bernabei:2010mq, Bernabei:2013xsa, Aalseth:2010vx,
  Aalseth:2011wp, Aalseth:2012if, Aalseth:2014eft,
Angloher:2011uu, Agnese:2013rvf}, an interpretation
which however, typically is in disagreement with bounds from other
experiments \cite{Ahmed:2009zw, Ahmed:2010wy, Angle:2011th,
  Aprile:2012nq, Kim:2012rza, Akerib:2013tjd, Agnese:2014aze}. Such a
conclusion depends on the assumed particle physics model for the
DM--nucleus interaction as well as on assumptions for the local DM
density and velocity distribution.

In this paper we are going to address the second issue by adopting
methods for comparing different experiments which are independent of
the assumed DM distribution. Our work is based on the so-called
minimal velocity ($v_m$) method proposed in Refs.~\cite{Fox:2010bz,
  Fox:2010bu}, which has been applied by a number of authors, see
e.g., \cite{McCabe:2011sr, Frandsen:2011gi, Gondolo:2012rs,
  HerreroGarcia:2012fu, Bozorgnia:2013hsa, DelNobile:2013cva,
  DelNobile:2013cta, DelNobile:2013gba, Frandsen:2013cna, Fox:2014kua,
  Feldstein:2014gza, Cherry:2014wia}. Usually this method is used to
derive bounds on the halo integral in $v_m$ space from data setting
limits, which then can be compared to the positive results from
experiments reporting a signal. This allows a qualitative assessment
whether a certain signal is in agreement or disagreement with bounds,
whereas a quantitative statement on the consistency is lacking. The
aim of this work is to present a way to quantify the compatibility of
a positive signal with limits from other experiments, extending
methods used in Ref.~\cite{HerreroGarcia:2012fu}.  We are going to
calculate an upper bound on the probability for both experimental
outcomes (the one reporting a signal and the one setting an upper
limit) to occur simultaneously, assuming the DM hypothesis.

Below we are going to consider data from the CDMS-II experiment using
a silicon target~\cite{Agnese:2013rvf} as well as the signal for
annual modulation from DAMA~\cite{Bernabei:2013xsa}. We calculate the
joint probability to obtain those indications in favour of DM
scattering together with the results from the
LUX~\cite{Akerib:2013tjd} and SuperCDMS~\cite{Agnese:2014aze}
experiments, which place strong limits using xenon and germanium
targets, respectively. In section~\ref{sec:data} we briefly describe
the data and our analysis thereof. After setting up the notation in
section~\ref{sec:notation}, we review the $v_m$ method to set upper
bounds on the halo integral in section~\ref{sec:bound}. In
section~\ref{sec:method} we present our method to calculate the joint
probability for observing a DM signal together with the data leading
to strong limits on the scattering cross section.  We consider the
situation encountered in CDMS silicon data, of an excess of few events
above the expected background in section~\ref{sec:method-events},
where we also introduce the ``signal length'' method. It provides a simple
way to calculate the probability of the experimental outcome taking into
account energy information of the obtained events. In section~\ref{sec:method-modulation} we apply our
method also to the case of DAMA observing a signal for annual
modulation, both using a ``trivial bound" on the modulation amplitude
as well as a bound based on the expansion of the halo integral in the
Earth's velocity~\cite{HerreroGarcia:2011aa, HerreroGarcia:2012fu}. 
In section \ref{sec:isospin} we show the results of our method for
the case of isospin violating DM-nucleus interactions. 
A general discussion and conclusions follow in
section~\ref{sec:discussion}.

In the appendix we give some details on the signal length method,
which allows us to calculate the probability of obtaining a given
experimental outcome taking into account the total number of events as
well as their energy distribution. It is inspired by the maximum-gap
method to set an upper limit \cite{Yellin:2002xd}; for a recent review
of statistical methods in astroparticle physics see
Ref.~\cite{Conrad:2014nna}. In App.~\ref{app:prob} we derive the
equation for the relevant probability needed in
section~\ref{sec:method-events}. In App.~\ref{app:CDMS} we discuss
some properties of the signal length test and apply it to CDMS silicon
data, showing that it leads to results consistent with the standard
likelihood analysis. A general discussion of the signal length test
follows in App.~\ref{app:discussion}.

During the preparation of this paper, the preprint
\cite{Feldstein:2014ufa} by Feldstein and Kahlhoefer appeared,
addressing a similar question. The method of
Ref.~\cite{Feldstein:2014ufa} uses a test statistic based on the joint
likelihood, whose distribution has to be calculated by Monte Carlo
simulation. Our approach is complementary to theirs and does not
require simulations. The distributions of the relevant statistics are
derived analytically from Poisson or Normal distributions.

\section{Description of the data used in this analysis}
\label{sec:data}

In order to illustrate our method with specific examples, we will use
in the following the positive signals from
DAMA/LIBRA~\cite{Bernabei:2013xsa} (DAMA for short) and the CDMS-II
silicon data~\cite{Agnese:2013rvf} (CDMS-Si for short) and compare
them to the limits from the LUX~\cite{Akerib:2013tjd} and
SuperCDMS~\cite{Agnese:2014aze} experiments. Based on this selection
of data sets we will demonstrate how to apply our method in the case
of a positive signal consisting of few events (CDMS-Si) as well as
annual modulation (DAMA).\footnote{We will not consider previous hints
  from the CoGeNT \cite{Aalseth:2010vx, Aalseth:2011wp,
    Aalseth:2012if, Aalseth:2014eft} and CRESST \cite{Angloher:2011uu}
  experiments, which most likely have a non-DM interpretation, see
  \cite{Aalseth:2014jpa, Davis:2014bla, kelso:2014-APP} and
  \cite{Angloher:2014myn}, respectively.} We proceed by giving a brief
description of the used data and our analysis thereof.

\bigskip

The {\bf LUX} (Large Underground Xenon) experiment has released its first
results~\cite{Akerib:2013tjd}. In their analysis of 85.3 live-days of
data taken in the period of April to August 2013, the data is
consistent with the background-only hypothesis. We consider as signal
region the region below the mean of the Gaussian fit to the nuclear
recoil calibration events (red solid curve in Fig.~4 of
\cite{Akerib:2013tjd}) and assume an acceptance of 0.5. It can be seen
from Fig.~4 of \cite{Akerib:2013tjd} that one event at 3.1
photoelectrons falls on the red solid curve. In this analysis we
assume zero events make the cut. Assuming the Standard Halo Model with
the Maxwellian velocity distribution and parameters chosen as in
\cite{Akerib:2013tjd}, we find that our 90\% CL contour agrees with
good accuracy with the limit set by the LUX collaboration. To find the
relation between S1 and nuclear recoil energy $E_R$, we use Fig.~4 of
\cite{Akerib:2013tjd} and find the value of S1 at the intersection of
the mean nuclear recoil curve and each recoil energy contour. For the
efficiency as a function of recoil energy, we interpolate the black
points in Fig.~9 of \cite{Akerib:2013tjd} for events with a corrected
S1 between 2 and 30 photoelectrons and a S2 signal larger than 200
photoelectrons. We multiply the efficiency from Fig.~9 of
\cite{Akerib:2013tjd} by 0.5 to find the total efficiency, and set it
equal to zero below $E_R=3$ keV.

\bigskip

The {\bf SuperCDMS} collaboration has observed eleven events in the recoil
energy range of [1.6, 10] keV with an exposure of 577 kg day of data
taken with their Ge detectors between October 2012 and June
2013~\cite{Agnese:2014aze}. The collaboration sets an upper limit on
the spin-independent WIMP-nucleon cross section of $1.2 \times
10^{-42}$~cm$^2$ at a WIMP mass of 8 GeV. For the detection efficiency
as a function of recoil energy, we use the red curve in Fig.~1 of
\cite{Agnese:2014aze}, and assume an energy resolution of 0.2 keV.

\bigskip

The {\bf DAMA} experiment has observed a 9.3~$\sigma$ annual
modulation signal during 14 annual cycles. We use the data on the
modulation amplitude for the total cumulative exposure of 1.33 ton yr
of DAMA/LIBRA-phase1 and DAMA/NaI given in Fig.~8 of
Ref.~\cite{Bernabei:2013xsa}. We consider small DM masses ($\le 20$
GeV) in this analysis, and can thus assume that the DAMA signal is
entirely due to scattering on Na.  For the quenching factor of Na we
take $q_{\rm Na}=0.3$ as measured by the DAMA
collaboration~\cite{Bernabei:1996vj}.

\bigskip

{\bf CDMS-Si} has observed three DM candidate events with recoil
energies of 8.2, 9.5, and 12.3 keV in their data taken with Si
detectors with an exposure of 140.2 kg day between July 2007 and
September 2008~\cite{Agnese:2013rvf}. The total estimated background
was 0.62 events in the recoil energy range of $[7,100]$ keV. To
include the background, we rescale the individual background spectra
from Ref.~\cite{McCarthy}, such that 0.41, 0.13, and 0.08 events are
expected from surface events, neutrons, and $^{206}$Pb,
respectively. We use the detector acceptance from
Ref.~\cite{Agnese:2013rvf} and assume an energy resolution of 0.3 keV.

\section{Notation}
\label{sec:notation}

We consider the case of elastic scattering of DM $\chi$ off a nucleus $(A,Z)$, depositing the nuclear recoil energy
$E_{nr}$ in the detector. The differential rate in events/keV/kg/day is given by
\beq \label{rate}
{R}(E_{nr},t) = \frac{\rho_\chi}{m_\chi} \frac{1}{m_A}\int_{v>v_{m}}d^3 v \frac{d\sigma_A}{d{E_{nr}}} v f_{\rm det}(\vect v, t),
\eeq
where $\rho_\chi \simeq 0.3 \, {\rm GeV/cm}^3$ is the local DM density, $m_A$
and $m_\chi$ are the nucleus and DM masses, $\sigma_A$ the DM--nucleus
scattering cross section, and $\vect v$ the 3-vector relative velocity
between DM and the nucleus, while $v\equiv |\vect{v}|$. The minimal velocity $v_{m}$ for a DM particle to
deposit a recoil energy $E_{nr}$ in the detector is
\beq
v_m=\sqrt{\frac{m_A E_{nr}}{2 \mu_{\chi A}^2}},
\label{vmin}
\eeq
where $\mu_{\chi A}$ is the reduced mass of the DM-nucleus system. 

For the standard  spin-independent and spin-dependent scattering  the differential cross section is 
\begin{align}
  \frac{d\sigma_A}{dE_{nr}} = \frac{m_A}{2\mu_{\chi A}^2 v^2} \sigma_A^0 F^2(E_{nr}) \,, 
  \label{eq:dsigmadE}
\end{align}
where $\sigma_A^0$ is the total DM--nucleus scattering cross section
at zero momentum transfer, and $F(E_{nr})$ is a form factor. We focus
here on spin-independent elastic scattering, where $\sigma_A^0$ can be written as
\beq
\sigma_A^0 = \sigma_p \left[ Z + (A - Z) \left(\frac{f_n}{f_p} \right) \right]^2 \left( \frac{\mu_{\chi A}}{\mu_{\chi p}} \right)^2,
\label{eq:sigma_A}
\eeq
where $\sigma_p$ is the DM-proton cross section, $f_{n,p}$ are coupling strengths to neutron and proton, respectively, and 
$\mu_{\chi p}$ is the reduced mass of the DM-nucleon system. In sections \ref{sec:bound} and \ref{sec:method} we assume  that DM couples with the same strength to protons and neutrons ($f_p=f_n$). We relax this assumption in section \ref{sec:isospin}, where we consider favourable choices of $f_n/f_p$.

One can define the halo integral as
\beq\label{eq:eta} 
\eta(v_m, t) \equiv \int_{v > v_m} d^3 v \frac{f_{\rm det}(\vect{v}, t)}{v} \,,
\eeq
where $f_{\rm det}(\vect{v}, t)$ is the DM velocity distribution in the detector rest frame.  Then the event rate can be written as
\beq\label{eq:Rate}
R(E_{nr}, t) = \frac{A^2 \, \sigma_p \, \rho_\chi }{2 m_\chi \mu_{\chi p}^2} \, F^2(E_{nr}) \, \eta(v_m, t) .
\eeq
The halo integral $\eta(v_m,t)$ parametrizes the astrophysics dependence of the event rate.

The DM velocity distribution in the detector rest frame is related to
the distribution in the rest frame of the Sun, $f(\vect{v})$, by
$f_{\rm det}(\vv, t) = f(\vv + \vv_e(t))$, where $\vv_e(t)$ is the
Earth's velocity around the Sun, with $v_e=29.8$~km/s. Since $v_e$ is
small compared to the velocity of the Sun with respect to the center
of the Galaxy ($v_{\rm Sun} \simeq 230$ km/s), one can expand the halo
integral Eq.~\eqref{eq:eta} in powers of $v_e$,
\begin{equation}
  \eta(v_m, t) = \bar \eta(v_m) + A_\eta(v_m) \cos 2\pi[t - t_0(v_m)] + \mathcal{O}(v_e^2).
\end{equation}
The zeroth order term, $\bar \eta(v_m)$, is responsible for the
unmodulated (time averaged) rate up to terms of order $v_e^2$. The
first order terms in $v_e$ lead to the annual modulation signal, with
$A_\eta(v_m)$ the amplitude of the annual modulation.

Let us define 
\beq
\tilde \eta(v_m)  \equiv \frac{\sigma_p \, \rho_\chi }{2 m_\chi \mu_{\chi p}^2} \, \bar \eta(v_m),
\eeq
with units of events/kg/day/keV. Then the predicted number of events in a detected energy interval $[E_1, E_2]$ can be written as 
\begin{equation}\label{eq:Npred}
 N^{\rm pred}_{[E_1,E_2]} = M T A^2 \int_0^\infty d E_{nr} F^2 (E_{nr}) G_{[E_1, E_2]} (E_{nr}) \tilde \eta(v_m),
\end{equation}
where $MT$ is the exposure of the experiment in units of kg~day, and $G_{[E_1,E_2]} (E_{nr})$ is the detector response function which
includes the detection efficiencies and energy resolution.

\section{Upper bound on the halo integral $\tilde{\eta}(v_m)$}
\label{sec:bound}

To obtain an upper bound on the unmodulated halo integral $\tilde{\eta}(v_m)$, 
we use the method discussed in Ref.~\cite{HerreroGarcia:2012fu} (see also
\cite{Fox:2010bz}).
Using the fact that $\tilde{\eta}(v_m)$ is a
falling function, the minimal number of events is obtained
for $\tilde{\eta}$ constant and equal to $\tilde{\eta}(v_m)$ up to
$v_m$ and zero for larger values of $v_m$. Therefore, for a given
$v_m$ we have a lower bound on the predicted number of events, $N^{\rm pred}_{[E_1,E_2]}$, in an
interval of observed energies $[E_1, E_2]$ of 
\beq
N^{\rm pred}_{[E_1,E_2]} > 
\mu(v_m) = MT A^2 \tilde{\eta}(v_m) \int_{0}^{E (v_m)} dE_{nr} F_A^2(E_{nr}) G_{[E_1,E_2]}(E_{nr})\,,
\label{eq:muvm}
\eeq
where the upper integration boundary $E(v_m)$ is given by Eq.~\eqref{vmin}.

Assuming an experiment observes $N^{\rm obs}_{[E_1,E_2]}$ events in
the interval $[E_1, E_2]$, we can obtain an upper bound on
$\tilde{\eta}(v_m)$ for a fixed $v_m$ at a confidence level CL by
requiring that the probability of obtaining $N^{\rm obs}_{[E_1,E_2]}$
events or less for a Poisson mean of $\mu(v_m)$ is equal to
$1-$CL. Let $\mu_{\rm CL}$ be the solution of the following equation for $\mu$:
\beq
e^{-\mu} \, \sum_{n=0}^{N^{\rm obs}} \frac{\mu^n}{n!} = 1 - {\rm CL} \,.
\label{eq:etabnd}
\eeq
Then an upper bound on $\tilde{\eta}(v_m)$ at the confidence level CL 
is obtained from Eq.~\eqref{eq:muvm} as
\beq\label{eq:eta-bnd}
\tilde{\eta}_{\rm bnd}(v_m) 
 =\frac{\mu_{\rm CL}}{MT A^2 \int_{0}^{E (v_m)} dE_{nr} F_A^2(E_{nr}) G_{[E_1,E_2]}(E_{nr})} \,.
\eeq
For LUX, $N^{\rm obs} = 0$ and Eq.~\eqref{eq:etabnd} just gives $\mu_{\rm CL} = - \log(1-{\rm CL})$. For SuperCDMS we have $N^{\rm obs} = 11$
and Eq.~\eqref{eq:etabnd} is solved numerically. Note that we make no assumptions about backgrounds and effectively assume that there is no background, which provides the most conservative limit on the DM signal. 
Note also that we do not bin the data for LUX and SuperCDMS but just require that the DM signal does not predict more events than observed in the total energy range.\footnote{In some cases, binning may lead to even stronger limits \cite{HerreroGarcia:2012fu}.}

This bound can now be compared to the results of other experiments,
seeing a positive signal for DM in the following way.  Suppose an
experiment observes an excess of events above their expected
background.  Let $N_i^{\rm obs}$ be the number of observed events in
the {\it i}'th recoil energy bin, and $\beta_i$ the expected
background in that bin.  Then we can use Eq.~\eqref{eq:Npred} to
experimentally determine the value of the halo integral in a given
bin:
\beq
\langle \tilde{\eta} (v_m^i) \rangle =\frac{N_{i}^{\rm obs} - \beta_i}{MT A^2 \int_{0}^{E (v_m)} dE_{nr} F_A^2(E_{nr}) G_i(E_{nr})} \,.
\label{eq:avg-eta}
\eeq
If the DM interpretation (and the background estimate) is correct this value should
satisfy 
 \beq \label{eq:avgeta-bnd}
 \langle \tilde{\eta} (v_m) \rangle \le \tilde \eta_{\rm bnd} (v_m) \,,
 \eeq
where the bound on the right-hand side is obtained from another
 experiment setting a limit via Eq.~\eqref{eq:eta-bnd}.

For the case in which an experiment observes an annual modulation signal, we can use 
a ``trivial bound'' on the modulation amplitude, which is based on the simple fact that the amplitude
of the first harmonic has to be smaller than the time averaged part, i.e., $A_\eta \le \bar \eta$, which is valid for any positive function. 
In case of a multi-target experiment observing an annual modulation signal (such as DAMA), we can assume that for a certain WIMP mass range the modulation signal is entirely due to scattering on one target nucleus T. Then in each energy bin in which there is a modulation signal we can write \cite{HerreroGarcia:2012fu}
\beq\label{eq:A-tilde}
\tilde{A}_\eta^{\rm obs} (v_m^i)=\frac{A_{i}^{\rm obs} q_{\rm T}}{A_{\rm T}^2 
\langle F_{\rm T}^2 \rangle_i f_{\rm T}} \,,
\eeq
where the index $i$ labels energy bins, $q_{\rm T}$ is the quenching
factor for target nucleus T, $\langle F_{\rm T}^2 \rangle_i$ is the
Helm form factor for target T averaged over the bin width, and $f_{\rm
  T}$ is the mass fraction of T in the multi-target experiment.  The
trivial bound applies to $\tilde \eta$ and $\tilde A_\eta$ without
change, thus for a fixed $v_m$ we have
\beq \label{eq:trivial}
\tilde{A}_\eta^{\rm obs} (v_m) \le \tilde \eta_{\rm bnd} (v_m) \,.
\eeq
\begin{figure}
\begin{center}
 \includegraphics[width=0.65\textwidth]{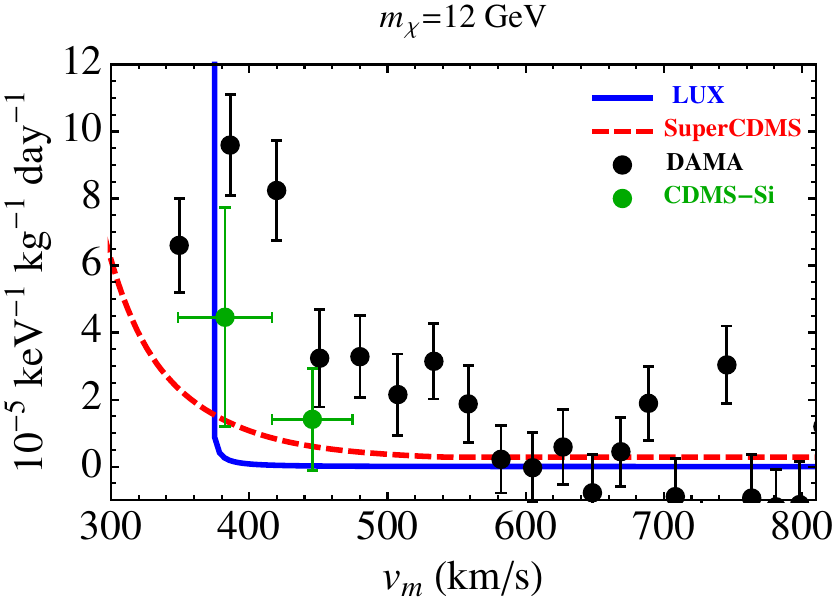}
\caption{\label{fig:eta} Upper bounds $\tilde \eta_{\rm bnd}$ at 3$\sigma$ from LUX and SuperCDMS are shown as solid blue and dashed red curves, respectively. The modulation amplitude $\tilde{A}_\eta^{\rm obs}$ for DAMA (Eq.~\eqref{eq:A-tilde}), and $\langle \tilde{\eta} \rangle$ derived from the excess of events observed by CDMS-Si (Eq.~\eqref{eq:avg-eta}) are shown as black and green points, respectively. Horizontal error bars for CDMS-Si indicate the adopted binning.
Spin-independent interactions with $f_n = f_p$ and a DM mass of 12 GeV are assumed.}
\end{center}
\end{figure}

In Fig.~\ref{fig:eta} we show an example of using the upper bound 
$\tilde{\eta}_{\rm bnd}(v_m)$ obtained from null results of the
LUX~\cite{Akerib:2013tjd} and SuperCDMS~\cite{Agnese:2014aze}
experiments to constrain the excess of events observed by
CDMS-Si~\cite{Agnese:2013rvf}, and the annual modulation signal
observed by the DAMA experiment~\cite{Bernabei:2013xsa}. 
Here we assume spin-independent interactions and a DM mass of 12~GeV.

The black data points and error bars in Fig.~\ref{fig:eta} show
$\tilde{A}_\eta^{\rm obs}$ and its corresponding error bars for DAMA
obtained from Eq.~\eqref{eq:A-tilde}. For CDMS-Si we consider two
energy bins of 3~keV width in the range of $[7,10]$ and $[10,13]$~keV,
containing two and one observed events, respectively. The green data
points and error bars in Fig.~\ref{fig:eta} show $\langle \tilde{\eta}
\rangle$ and its corresponding error bars for CDMS-Si obtained from
Eq.~\eqref{eq:avg-eta} for the two bins.\footnote{Binning is used only for the purpose of showing the data in Fig.~\ref{fig:eta}. For the probability analysis of CDMS-Si data given below no binning is required.} It is clear from
Fig.~\ref{fig:eta} that the upper bounds from LUX and SuperCDMS are in
tension with the modulation signal from DAMA and to some extent also
with the excess of events observed in CDMS-Si for
$m_\chi=12$~GeV. While in the case of DAMA the situation is rather
clear, indicating conflict at very high CL, for CDMS-Si a more
quantitative way of reporting agreement or disagreement is needed. In
the following we will provide methods for this purpose.

\section{Joint probability of positive and negative results}
\label{sec:method}

\subsection{Experiments observing excess of events}
\label{sec:method-events}

In this subsection we focus on the case in which an experiment
observes an excess of events above their background, such as the
case of CDMS-Si. We provide two methods for quantifying the
disagreement between the observed excess of events by experiment A and
the rate from null-result experiment B, one using only total event
numbers and the other using in addition the energy information.

Let us define $p_B$ as the probability to obtain equal or less events
than observed by the null-result experiment B for a Poisson mean of
$\mu(v_m)$ as defined in Eq.~\eqref{eq:etabnd}, with $p_B = 1 -
$CL. Then one can use the upper bound $\tilde{\eta}_{\rm bnd}^B(v_m)$,
Eq.~\eqref{eq:eta-bnd}, obtained at a confidence level $1- p_B$ from
experiment B in Eq.~\eqref{eq:Npred} to get an upper bound on the
predicted number of events in experiment A,
\beq
 N^{{\rm bnd},A}_{[E_1,E_2]} = M T A^2 \int_0^\infty d E_{nr} F^2 (E_{nr}) G_{[E_1, E_2]} (E_{nr}) \tilde \eta_{\rm bnd}^B(v_m).
 \label{eq: Nbnd}
\eeq
In order to compute the integral on the r.h.s.~of Eq.~\eqref{eq: Nbnd}, $\tilde \eta_{\rm bnd}^B(v_m)$ has to be written as a function of the recoil energy deposited in the detector of experiment A.

In practice, one also has to include the expected number of background events in the upper bound on the predicted number of events. If $\beta_{[E_1,E_2]}^A$ is the number of background events expected by experiment A in the energy interval $[E_1, E_2]$, then we have an upper bound on the predicted number of events $N^{{\rm pred},A}_{[E_1,E_2]} \leq \mu_{\rm bnd}^A$, where 
\beq \label{eq:muA}
\mu_{\rm bnd}^A = N^{{\rm bnd},A}_{[E_1,E_2]} + \beta_{[E_1,E_2]}^A \,. 
\eeq
Note that $N^{{\rm bnd},A}_{[E_1,E_2]}$ depends on the CL that
$\tilde{\eta}_{\rm bnd}^B$ is obtained at, and thus it depends on
$p_B$. In this work we always assume that the expected background is
known. The motivation for this is that we are interested in the situation
of very few signal events (such as in CDMS-Si) and in this case the
statistical errors are larger than the assumed uncertainty in the
background.

\subsubsection{Method 1 -- total number of events}

In the first method we only use the information on the total number of observed events in the full energy interval. The probability $p_A$ of obtaining $N^{{\rm obs},A}_{[E_1,E_2]}$ events or more by experiment~A for a Poisson mean of $\mu_{\rm bnd}^A$ is given by
\beq
p_A = e^{-\mu_{\rm bnd}^A} \, \sum_{n=N^{{\rm obs},A}}^{\infty} \frac{\left(\mu_{\rm bnd}^A \right)^n}{n!} \,.
\eeq
The combined probability of obtaining the results of experiment A
and experiment B is given by $p_A ~ p_B$, where $p_A$ is a function 
of the chosen $p_B$, see Eqs.~\eqref{eq: Nbnd} and \eqref{eq:muA}.
We can then calculate the largest possible joint
probability by maximizing with respect to $p_B$:
\beq\label{eq:pjoint1}
p_{\rm joint} = \max_{p_B}\left[ p_A(p_B) \,\, p_B \right] \,.
\eeq

When applying this method some care has to be taken when choosing the energy intervals of the two experiments. We need to make sure that experiment B provides a limit over the full energy range considered in Eq.~\eqref{eq: Nbnd} for experiment A. According to Eq.~\eqref{vmin}, the recoil energies in experiments A and B probing the same $v_m$ are related by
\be\label{eq:Erelation}
\frac{E_A}{E_B} = \frac{m_B \, \mu^2_{\chi A}}{m_A \, \mu^2_{\chi B}} \,.
\ee
If the lower edge of the interval for experiment A, $E_1$ in
Eq.~\eqref{eq: Nbnd}, is below the threshold of experiment B (after being
translated into experiment A energies according to
Eq.~\eqref{eq:Erelation}), no limit can be obtained for the expected
number of events for experiment A since $\tilde \eta$ is unbounded by
experiment B below its threshold. Due to the finite energy resolution
of experiment A (included in the function $G_{[E_1,E_2]}(E_{nr})$ in
Eq.~\eqref{eq: Nbnd}) a bound on $\tilde \eta$ is even required below
the reconstructed energy $E_1$. In our analysis of CDMS-Si we require
that $E_1$ is equal to (or larger than) the thresholds of SuperCDMS or LUX translated
into Si recoil energies according to Eq.~\eqref{eq:Erelation} plus 3 times
the energy resolution of CDMS-Si.

\begin{figure}
\begin{center}
 \includegraphics[width=0.65\textwidth]{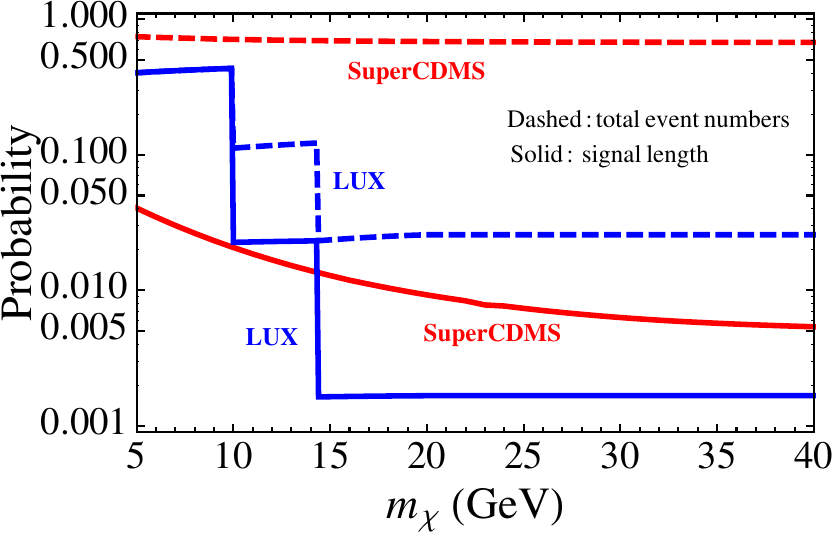}
\caption{\label{fig:prob-Si} The probability that the excess of events
  observed by CDMS-Si is compatible with null-results from LUX (blue)
  and SuperCDMS (red) as a function of WIMP mass. The dashed and solid
  curves correspond to Method 1 (Poisson probabilities of total number
  of events in the full energy range) and Method 2 (including energy
  information of the events via the ``signal length'') of obtaining the
  joint probability, respectively.}
\end{center}
\end{figure}

In Fig.~\ref{fig:prob-Si} we show the probability that the excess of
events observed by CDMS-Si is compatible with null-results from LUX
(dashed blue) and SuperCDMS (dashed red) as a function of WIMP mass.
For SuperCDMS we find that the threshold of 1.6~keV corresponds to Si
energies which are always smaller than the 7~keV threshold of CDMS-Si
(including the energy resolution) for the whole range of DM masses
shown in the figure. Hence, in this case we consider the full recoil
energy range of $[7,100]$~keV for CDMS-Si. For LUX, however, the
threshold of the efficiency at 3~keV translated into Si energies plus
3 times the energy resolution is larger than the 7~keV CDMS-Si threshold for
DM masses $m_\chi \lesssim 20$~GeV. In this case we set $E_1$ in Eq.~\eqref{eq: Nbnd} to
\be\label{eq:E1}
E_1 = \frac{m_{\rm Xe} \, \mu^2_{\chi \rm Si}}{m_{\rm Si} \, \mu^2_{\chi \rm Xe}} (3 \,{\rm keV}) + 3\sigma_{\rm res} \,,
\ee
where $\sigma_{\rm res} = 0.3$~keV is the energy resolution we assume
for CDMS-Si.  Then, if $E_1$ computed in this way is larger than the
energy of the first, second, or third event observed in CDMS-Si, we
take 2, 1, or 0 observed events in the modified energy range,
respectively. This is responsible for the steps observed in the LUX
curves in Fig.~\ref{fig:prob-Si}: in the regions $m_\chi <
10\,{\rm GeV}$, $10\,{\rm GeV}  \leq  m_\chi < 14\,{\rm GeV}$, $m_\chi \geq
14\,{\rm GeV}$ there are 1, 2, 3 events in the analysis window,
respectively.

We see that based on this method the joint probability of CDMS-Si and
SuperCDMS is around 70\%, i.e., signaling essentially
compatibility. For CDMS-Si and LUX the joint probability for $m_\chi
\gtrsim 14$~GeV (where all 3 CDMS-Si events are included in the
analysis window) approaches the Poisson probability for the background-only
hypothesis, i.e., the probability to obtain 3 or more events for an
expectation of 0.62 background events, which is 2.57\%. In the
following we show that if some information on the energy of the
observed events is used, significantly smaller joint probabilities are
obtained.

\subsubsection{Method 2 -- the ``signal length'' method}
\label{sec:method-signal-length}

In the second method for computing the probability, we use the energy
information of the events in addition to the observed number of
events. Our method is inspired by the widely used maximum-gap method
to set an upper limit \cite{Yellin:2002xd}. Instead of considering the
gap between two events we find it useful to look at the 
``signal length'' (SL), defined in the following way. 

Consider an experiment observing
$n \ge 2$ events. Then we define the signal length $\Delta$ as
\beq
\begin{array}{rl}
\Delta \equiv & \text{expected number of events in the energy interval between}\\ 
              & \text{the two events with the lowest and highest energy.}
\end{array}
\eeq
The signal length can be calculated for a given DM model and halo, and includes
also the background expectation. Furthermore, let $\mu$ denote the total
number of expected events (including background) in the full energy
interval for the experiment. Clearly, we have $\Delta \le \mu$.
Suppose experiment A observes $N^{{\rm obs}}$ events.
For a given DM model, DM halo, and background, we can calculate 
the signal length $\Delta$ and the total number of expected events $\mu$ in experiment A. 
Consider now the probability
\be 
P(n \ge N^{\rm obs}, \, {\rm SL} \le \Delta | \mu) \equiv P_{\rm SL}(N^{{\rm obs}}, \Delta|\mu) \,,
\ee
i.e., the joint probability of obtaining $N^{{\rm obs}}$ or more
events and a signal length of size $\Delta$ or smaller.
In App.~\ref{app:prob} we derive the expression for this probability:
\begin{align}
  P_{\rm SL}(N^{{\rm obs}}, \Delta|\mu) &= e^{-\mu} \sum_{n=N^{{\rm obs}}}^\infty \frac{1}{n!} [n\mu \Delta^{n-1} - (n-1)\Delta^n] \,.
\label{eq:prob}
\end{align}
We will use this probability in order to quantify how likely an
observation of $N^{{\rm obs}}$ and $\Delta$ is to occur for given
$\mu$. It takes into account both the total number of events, as well
as some information on the energy distribution of the events.  The
motivation to consider the signal length is that generically a DM
signal is expected to be concentrated in a small energy interval
(typically at low energies), whereas background distributions are
often more extended. The signal length method is designed to discriminate a
signal predicting clustered events from a more broadly distributed
background. For instance, we can calculate the probability
Eq.~\eqref{eq:prob} for the background-only hypothesis in CDMS-Si.  In
this case, $\mu = 0.62$ events (the total background expectation) and
$\Delta = 0.104$ (integrating the background between 8.2 and
12.3~keV), and we find a probability of $P_{\rm SL} = 0.17\%$, which is
close to the probability of 0.19\% obtained by a likelihood-ratio test
between the DM and background only
hypotheses~\cite{Agnese:2013rvf}.\footnote{For a given DM halo one can
  use the probability Eq.~\eqref{eq:prob} also to test specific DM
  models. For instance, using CDMS-Si data and assuming the so-called
  standard halo model, contours of $P_{\rm SL}$ in the plane of $m_\chi$
  and $\sigma_p$ lead to regions which agree very well with the
  standard allowed regions based on a delta log-likelihood analysis,
  see Fig.~\ref{fig:SHM} in the appendix.}

Some comments are in order. Eq.~\eqref{eq:prob} is defined only for
$N^{\rm obs} \ge 2$.  If $\mu \lesssim 2$ the probability becomes
small because then it is unlikely to obtain 2 or more events. For $\mu
\gtrsim 2$ one has to compare $P_{\rm SL}(N^{{\rm obs}}, \Delta|\mu)$ with
the value for ``typical'' outcomes for $N^{{\rm obs}}$ and
$\Delta$. In App.~\ref{app:SL} we show that the expectation value of
$P_{\rm SL}$ for $\mu \gtrsim 2$ is close to 0.2. Hence we can conclude
that if for given $N^{{\rm obs}}$ and $\Delta$ the value of
$P_{\rm SL}(N^{{\rm obs}}, \Delta | \mu)$ is much smaller than 0.2 those
outcomes are quite unlikely, whereas values around 0.2 correspond to a
likely outcome. Further discussion of the signal length method is
given in the appendix.

\bigskip

In order to apply this method to our case of interest, we need to take
into account that we cannot predict $\mu$, but instead have only an
upper bound $\mu_{\rm bnd}$. Furthermore, we can calculate an upper
bound on the expected number of events in the energy interval between
the event with the lowest and highest energy: $\Delta_{\rm
  bnd}$. Additionally, there may be some background, assumed to be
known. Let us denote by $B$ the total number of expected background
events and by $b$ the expected background events in the interval
between the events with the lowest and highest energies. Hence, the
``true'' values of $\Delta$ and $\mu$ are bounded as 
\begin{align}
b &\le \Delta \le \Delta_{\rm bnd}, \\
\mu_{\rm lo}(\Delta) &\le \mu \le \mu_{\rm bnd}, \label{eq:mu}
\end{align}
with
    \begin{equation}\label{eq:mu_lo}
     \mu_{\rm lo}(\Delta) = B + (\Delta - b)\left[1 + 
      \frac{\int_0^\infty d E_{nr} F^2 (E_{nr}) G_{[E_1, E_1^\Delta]} (E_{nr})}
           {\int_0^\infty d E_{nr} F^2 (E_{nr}) G_{[E_1^\Delta, E_2^\Delta]} (E_{nr})} 
      \right]\,.
    \end{equation}
Here $[E_1, E_2]$ is the total energy interval, $[E_1^\Delta,
  E_2^\Delta]$ is the interval delimiting the signal length, i.e.,
$E_1^\Delta$ and $E_2^\Delta$ are the energies of the lowest and
highest events. In the lower bound for $\mu$ we have taken into
account that $\mu$ has to be larger than $\Delta$ plus the background
$B-b$ outside the $\Delta$-interval. The second term in the square
bracket of Eq.~\eqref{eq:mu_lo} follows from the fact that $\tilde
\eta (v_m)$ has to be a decreasing function. Then, for a given
$\Delta$ there is a lower bound on the expected event rate in the
interval $[E_1, E_1^\Delta]$ below the signal. Note that $\mu_{\rm
  bnd}$ and $\Delta_{\rm bnd}$ also include the number of expected
background events in the corresponding energy regions, and depend on
the CL.

We have to maximize the probability in Eq.~\eqref{eq:prob} with
respect to $\Delta$ and $\mu$, taking into account the allowed ranges
for them. First, it is easy to see that Eq.~\eqref{eq:prob} is a monotonously
increasing function in $\Delta$. Hence the maximum probability is
obtained by setting $\Delta = \Delta_{\rm bnd}$. Second, by differentiating
Eq.~\eqref{eq:prob} with respect to $\mu$ one finds that it has a maximum for
\begin{equation}
  \hat\mu = 1 + \frac{\sum_{n=N^{{\rm obs}}}^\infty \frac{n-1}{n!}\Delta^n}
                     {\sum_{n=N^{{\rm obs}}}^\infty \frac{1}{(n-1)!}\Delta^{n-1}} \,.
\end{equation}
Now we have to take into account the allowed range for $\mu$ in
Eq.~\eqref{eq:mu}. Hence, in order to maximize the probability we
define:\footnote{For the analyses reported in the following it turns
  out that the first case in Eq.~\eqref{eq:mu_max} never applies. For
  SuperCDMS versus CDMS-Si we are in the second case of
  Eq.~\eqref{eq:mu_max}, while for LUX versus CDMS-Si always the third
  case applies.  Hence, the precise value of the lower bound in
  Eq.~\eqref{eq:mu_lo} is not important. In particular the second term
  in the square bracket of Eq.~\eqref{eq:mu_lo} is always small and
  does not contribute to the result.}
\begin{equation}\label{eq:mu_max}
  \mu_{\rm max} \equiv \left\{ 
  \begin{array}{l@{\quad\text{for}\quad}l}
    \mu_{\rm lo}(\Delta_{\rm bnd}) & \hat\mu < \mu_{\rm lo}(\Delta_{\rm bnd}) \\
    \hat\mu &   \mu_{\rm lo}(\Delta_{\rm bnd}) \le \hat\mu \le \mu_{\rm bnd} \\
    \mu_{\rm bnd} & \hat\mu > \mu_{\rm bnd}
  \end{array} \right.
\end{equation}
Now we can use $P_{\rm SL}(N^{{\rm obs}},\Delta_{\rm bnd} | \mu_{\rm max})$
to check how likely a given outcome is even if only an upper bound on
the event rate is available.

In order to use the signal length probability to evaluate the joint
probability of experiment A (seeing a signal) and experiment B (giving
a limit) we proceed as follows. We specify the probability for
experiment B, $p_B$, and calculate an upper bound on the halo integral
$\tilde{\eta}_{\rm bnd}^B(v_m)$ from Eq.~\eqref{eq:eta-bnd} at the
confidence level CL~$= 1- p_B$. This bound is then used in
Eqs.~\eqref{eq: Nbnd} and \eqref{eq:muA} to calculate the upper bounds
$\mu_{\rm bnd}$ and $\Delta_{\rm bnd}$ for experiment A, by using in
those equations for the energy interval $[E_1, E_2]$ either the total
energy interval of the experiment (for $\mu_{\rm bnd}$) or the energy
interval between the events with smallest and largest energies (for
$\Delta_{\rm bnd}$). Then we can calculate $P_{\rm SL}(N^{{\rm
    obs}},\Delta_{\rm bnd} | \mu_{\rm max})$ for that particular
choice of $p_B$. As before, we maximize with respect to $p_B$
to obtain the maximal joint probability for the combined result:
\be\label{eq:pjoint2}
p_{\rm joint}= \max_{p_B} \left[P_{\rm SL}(N^{{\rm obs}},\Delta_{\rm bnd} | \mu_{\rm max}) ~ p_B\right] \,.
\ee

The solid blue and red curves in Fig.~\ref{fig:prob-Si} show the
results of such an analysis for the compatibility between the signal
in CDMS-Si and the null-results of LUX and SuperCDMS, respectively. We
find significantly smaller probabilities than in the case of using
total event numbers only (dashed curves), illustrating the importance
of using energy information and the power of the signal length
test. The joint probability of CDMS-Si and SuperCDMS is 4\% for
$m_\chi \simeq 5$~GeV, decreasing to 0.5\% for $m_\chi \simeq
40$~GeV. For LUX we proceed as before, taking as energy threshold the
maximum of the threshold energies of the two experiments, after
equalizing them via the $v_m$ method, see Eq.~\eqref{eq:E1}. The
step-like structure of the probability emerges from the fact that when
decreasing the DM mass, the Si-equivalent energy of the LUX threshold
is increasing. The steps occur when the threshold passes the energies
of the observed events. For $m_\chi \lesssim 10$~GeV only one event is
left in the analysis window and the signal-length method can no longer
be applied (since it is defined only for $\ge 2$ events). In this case
we use method~1, calculating just the Poisson probability to obtain one
event, which essentially provides no constraint given the expected
background. On the other hand, for $m_\chi \gtrsim
14$~GeV, when all three events are inside the analysis interval, we
find joint probabilities very close to the probability of the
background only hypothesis for CDMS-Si of 0.17\%, which is the maximal
possible rejection CL of the signal.\footnote{The fact that the LUX
  probabilities for method~1 above 14~GeV and method~2 below 14~GeV
  are similar seems to be a numerical accident. We have checked that
  artificially changing the expected background for CDMS-Si leads to
  different probabilities for those two cases.}

\subsection{Experiments observing annual modulation}
\label{sec:method-modulation}

Let us now show how to quantify the disagreement between an experiment
A observing an annual modulation signal and a null-result experiment B
providing an upper limit on the unmodulated rate
\cite{HerreroGarcia:2012fu}. We first consider the trivial bound given
in Eq.~\eqref{eq:trivial}, demanding that the amplitude of the
modulation is smaller than the bound on the unmodulated rate.  We
calculate the probability $p_B$ to obtain equal or less events than
observed by the null-result experiment B for each value of $\tilde
\eta_{\rm bnd}^B (v_m)$.  Using the same value of $\tilde \eta_{\rm
  bnd}^B (v_m)$ on the r.h.s.\ of Eq.~\eqref{eq:trivial}, we calculate
the probability $p_A$ to obtain a value of the modulation amplitude in
a fixed energy bin equal to or larger than the observed one, assuming a
Gaussian distribution for it, with a mean given by $\tilde \eta_{\rm
  bnd}^B (v_m)$ and a standard deviation given by the experimental
error on the modulation amplitude.  Hence, for a given energy bin $i$
we have
\beq
p_A = \frac{1}{2} \left [1 - \erf \left( \frac{\tilde A^{\rm obs}_\eta (v_m^i)-\tilde \eta_{\rm bnd}^B (v_m^i) }{\sqrt{2} \, \sigma_ {\tilde A}(v_m^i)} \right)  \right],
\label{eq: pA-mod}
\eeq 
where $\sigma_{\tilde A}(v_m^i)$ is the experimental error on $\tilde
A^{\rm obs}_\eta$ in bin $i$.\footnote{Note that this choice for the
  standard deviation of the Gaussian is conservative, since we take
  the experimental error on $\tilde A^{\rm obs}_\eta$ as an estimate
  for the standard deviation assuming a mean value $\tilde \eta_{\rm
    bnd}^B$. Since $\tilde \eta_{\rm bnd}^B$ is typically smaller than
  $\tilde A^{\rm obs}_\eta$ one would expect that the true statistical
  error is also smaller than the statistical error on $\tilde A^{\rm
    obs}_\eta$.}  The joint probability of obtaining the experimental
result for a fixed value of $\tilde \eta_{\rm bnd}^B$ is given by 
$p_A (p_B) ~ p_B$, and the highest possible joint probability is obtained 
by maximizing with respect to $p_B$:
\beq
p_{\rm joint} = \max_{p_B} \left[p_A(p_B) ~ p_B \right].
\eeq

\begin{figure}
\begin{center}
 \includegraphics[width=0.65\textwidth]{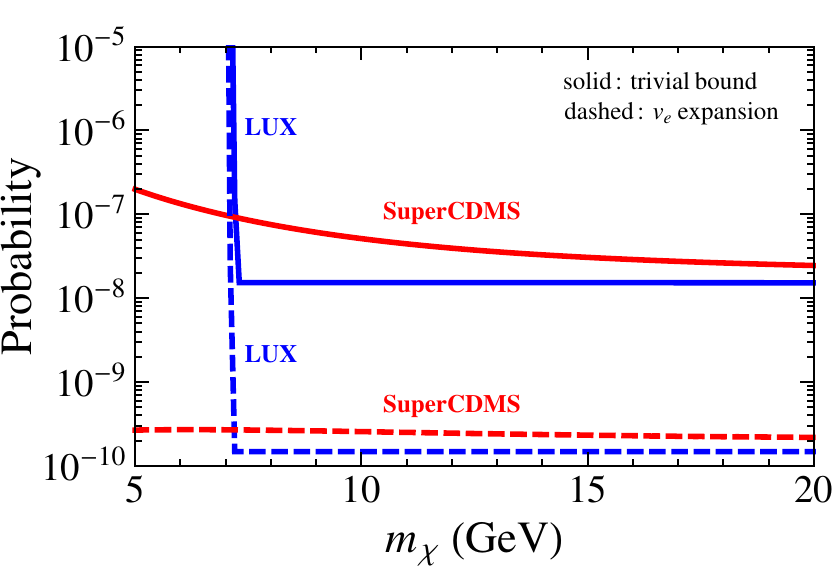}
\caption{\label{fig:DAMA} The probability that the modulation
  amplitude in DAMA is compatible with the null-results from LUX (blue
  curves) and SuperCDMS (red curves). The solid curves correspond to
  the trivial bound, requiring that the modulation amplitude is less
  than the bound on the unmodulated rate. $v_m$ is fixed at a value
  corresponding to the 3rd modulation data point in DAMA. The dashed
  curves correspond to the bound based on the expansion in $v_e$,
  Eq.~\eqref{eq:gen-bound}. The integration of the modulation
  amplitude starts from the 3rd bin in DAMA.}
\end{center}
\end{figure}

As an example, we apply our method to the case of the DAMA annual
modulation signal. We perform the analysis at a fixed $v_m$ which
corresponds to the center of the 3rd modulation data point in DAMA and depends on
the DM mass. This choice is arbitrary, but motivated by
Fig.~\ref{fig:eta}, which suggests strong tension around the $v_m$
corresponding to the 3rd bin. The probability that the DAMA modulation
amplitude in that bin is compatible with the constraints on $\tilde
\eta$ from LUX and SuperCDMS is shown by solid blue and red curves in
Fig.~\ref{fig:DAMA}, respectively, indicating very strong tension
between the results. From Fig.~\ref{fig:DAMA}, one can see a sharp
increase in the probability for LUX at low masses. Again this sharp
cut-off is due to setting the LUX detection efficiency equal to zero
below $E_{nr}=3$~keV. This results in having no upper bound on $\tilde
\eta$ at small DM masses ($\le 7$ GeV), when the minimal velocity
threshold (corresponding to 3~keV recoil energy) in LUX becomes larger
than the $v_m$ corresponding to the energy of the 3rd modulation data
point in DAMA.

In Refs.~\cite{HerreroGarcia:2011aa, HerreroGarcia:2012fu} it has been
shown that under certain regularity assumptions on the DM halo a much
stronger bound on the modulation amplitude can be obtained by
expanding the halo integral in the small Earth's velocity $v_e$.
Under the assumption that the DM velocity distribution in
the Sun's rest frame is constant in time on the scale of 1 year and in
space on the scale of the size of the Sun--Earth distance, the
modulation amplitude is bounded as
\begin{align}\label{eq:gen-bound}
 \int_{v_1}^{v_2} d v_m \tilde A_\eta^{\rm obs}(v_m) \le v_e 
  \left[ \tilde \eta_{\rm bnd}(v_1) + \int_{v_1}^{v_2} dv \frac{\tilde \eta_{\rm bnd} (v)}{v}  \right]\,.
\end{align}
We can compute the disagreement between the observed annual modulation signal in one experiment and the rate from another experiment in a way similar to that of the trivial bound, except that  for the bound in Eq.~\eqref{eq:gen-bound} we calculate $p_A$ by assuming a Gaussian distribution on the l.h.s.~of Eq.~\eqref{eq:gen-bound}, and for the integral on the r.h.s.~of Eq.~\eqref{eq:gen-bound} we take the bound
$\tilde \eta_{\rm bnd} (v_m)$ at constant probability $p_B$.

We apply our method for the bound in Eq.~\eqref{eq:gen-bound} to the
case of DAMA. For the lower and upper integration limits, $v_1$ and
$v_2$ in Eq.~\eqref{eq:gen-bound}, we take the $v_m$ corresponding to
the beginning of the 3rd and the end of the 12th bins in DAMA,
respectively. Above the 12th energy bin (i.e.~above 8 keVee), the data
is consistent with no modulation.  The dashed blue and red curves in
Fig.~\ref{fig:DAMA} show the probability that the integrated
modulation amplitude in DAMA is compatible with the bound (r.h.s.~of
Eq.~\eqref{eq:gen-bound}) derived from constraints on $\tilde \eta$
from LUX and SuperCDMS, respectively. As expected, the bound based on
the expansion in $v_e$ is a few orders of magnitude stronger than the
trivial bound, and thus the compatibility between the DAMA signal and
the results from LUX and SuperCDMS becomes weaker for the $v_e$
expansion bound compared to the trivial bound. A similar analysis has
been performed in Ref.~\cite{HerreroGarcia:2012fu} also for different
experiments and different DM--nucleus interactions, such as
spin-dependent interactions and interactions with arbitrary couplings
to protons and neutrons.

\section{Isospin violating interactions}
\label{sec:isospin}

In the previous sections we assumed the so-called isospin conserving DM-nucleon interaction, i.e.~that DM couples with the same strength to protons and neutrons, such that $f_n = f_p$. However, in general this assumption does not need to be satisfied, e.g.~\cite{Chang:2010yk, Feng:2011vu}. In particular, in the case of a relative sign between $f_n$ and $f_p$ there could be a suppression factor for the spin-independent DM-nucleus cross section (see Eq.~\eqref{eq:sigma_A}) due to a cancellation between the contributions from neutrons and protons, depending on the target element of the experiment. Since this can have important consequences for the compatibility of different experiments, in this section we consider a few interesting cases in which $f_n \neq f_p$.

\begin{figure}
\begin{center}
 \includegraphics[width=0.49\textwidth]{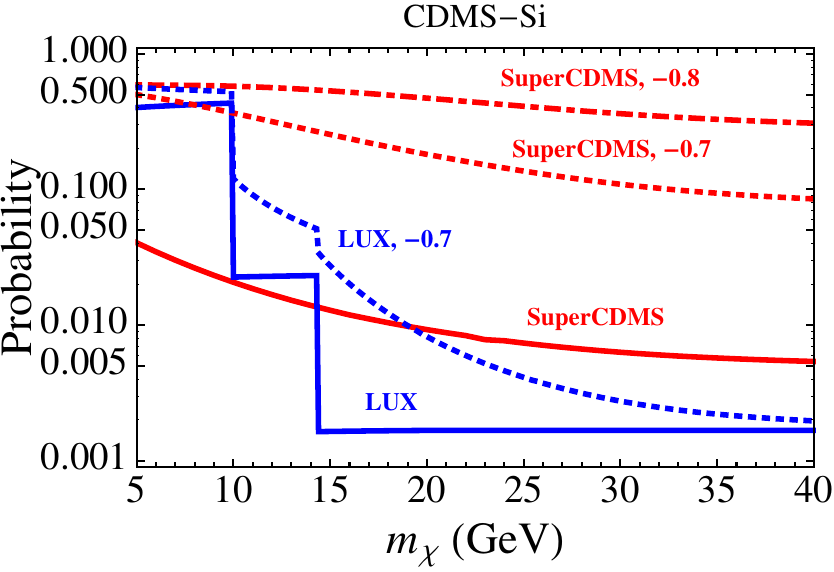}
  \includegraphics[width=0.49\textwidth]{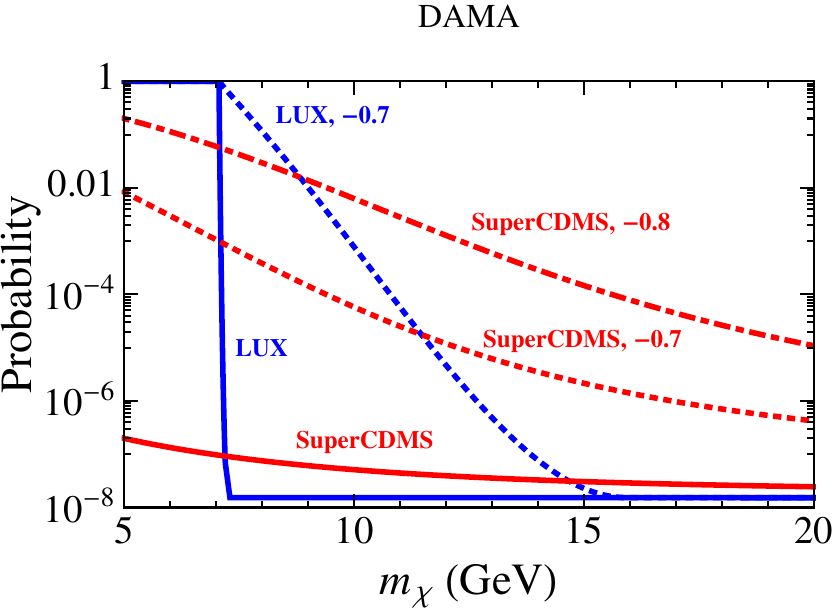}
\caption{\label{fig:IV} The probability that the excess of events
  observed by CDMS-Si using the signal length method (left panel) or
  the modulation amplitude in DAMA using the trivial bound at the 3rd
  modulation data point (right panel) is compatible with null-results
  from LUX (blue) and SuperCDMS (red). The solid curves
  correspond to isospin conserving interactions, whereas the dotted
  and dot-dashed curves correspond to isospin violating interactions
  with $f_n/f_p = -0.7$ and  $-0.8$, respectively. The LUX curve for 
  $f_n/f_p = -0.8$ coincides with the isospin conserving curve (solid).}
\end{center}
\end{figure}

We take into account the natural abundances of the isotopes present in each detector to compute the cross section, but neglect the effect of different isotopes on the form factors and kinematics. The cross section reaches a minimum at $f_n/f_p \simeq -0.7$ and $-0.8$ for Xe and Ge, respectively (see the left panel of Fig.~5 in \cite{Schwetz:2011xm})\footnote{We neglect here higher order QCD effects, which might be relevant in case of cancellations of the leading contributions to the cross section \cite{Cirigliano:2012pq, Cirigliano:2013zta}.}. Since for those values of $f_n/f_p$, the cross section is not highly suppressed for Si and Na, the compatibility between CDMS-Si/DAMA and LUX/SuperCDMS can be improved. A comparison of the joint probabilities for the isospin conserving and isospin violating cases is shown in Fig.~\ref{fig:IV}. In the left panel we show the joint probability of obtaining the positive CDMS-Si result and the negative LUX and SuperCDMS results, using the signal length method. In the right panel, the joint probability of obtaining the DAMA modulation signal and the null results from LUX and SuperCDMS is shown, using the trivial bound at a fixed $v_m$ corresponding to the 3rd modulation data point in DAMA. Clearly, assuming a value of $f_n/f_p$ which suppresses the cross section for Xe and Ge, leads to a higher joint probability. In the figure we assume $f_n/f_p = -0.7$ and $-0.8$, which maximally suppresses the cross section for Xe and Ge, respectively. For LUX, the cross section is not suppressed by much at $f_n/f_p = -0.8$ and the curves for that case coincide with the isospin conserving curves.

We find that the signal in CDMS-Si essentially becomes consistent with
SuperCDMS for $f_n/f_p = -0.8$, while it is inconsistent with LUX for
$m_\chi \gtrsim 14$ as in the isospin conserving case.  For $f_n/f_p =
-0.7$, the joint probability of CDMS-Si and SuperCDMS decreases to
18\% for $m_\chi \simeq 20$ GeV, and the joint probability with LUX
remains below 1\% for $m_\chi \geq 19$~GeV. For DAMA the compatibility
with LUX for $f_n/f_p = -0.7$ and with SuperCDMS for $f_n/f_p =
  -0.8$ is increased by many orders of magnitude for $m_\chi \lesssim
10$~GeV compared to the isospin conserving case ($p_{\rm joint}
\gtrsim 1\%$). However the compatibility of DAMA cannot be
  improved considerably with both LUX and SuperCDMS for a fixed choice
  of $f_n/f_p$. For example, for $f_n/f_p = -0.7$ the
  joint probability of DAMA and SuperCDMS always remains below 1\%.
Let us also note that for a given value of $f_n/f_p$, other data not
considered here may still be in considerable tension with DAMA.
Furthermore, using the bound based on the $v_e$ expansion
(Eq.~\eqref{eq:gen-bound}) stronger limits can be obtained also in the
isospin violating case.  For instance, in
Ref.~\cite{HerreroGarcia:2012fu} it was shown that for $f_n/f_p =
-0.7$ data from the CDMS silicon target \cite{Akerib:2005kh} leads to
joint probabilities with DAMA between $10^{-7}$ and $10^{-4}$ for DM
masses between 5 and 20~GeV.

\section{Summary and discussion}
\label{sec:discussion}

In the interpretation of results from DM direct detection experiments,
uncertainties related to the local DM distribution are crucial. The
so-called $v_m$ method allows for a completely halo independent
comparison of different experiments by considering constraints in
terms of the halo integral $\tilde \eta(v_m)$. Starting from this
idea, we have presented a method to evaluate the joint probability of
obtaining the outcomes of two potentially conflicting experiments,
under the assumption that the DM hypothesis is true. This allows a
quantitative assessment of the compatibility of such results.  

The main idea is the following. For a given value of $\tilde
\eta(v_m)$ we calculate the probabilities $p_A$ and $p_B$ of obtaining
the outcomes of the two experiments A and B, respectively. The joint
probability is just given by the product $(p_A p_B)$. Then we report 
the maximal possible joint compatibility by
maximizing this product with respect to
$\tilde\eta(v_m)$. For technical reasons it turns out to be practical
to consider $p_A$ as a function of $p_B$ and maximize $(p_A p_B)$
with respect to $p_B$, as described in detail in
section~\ref{sec:method}.

We have illustrated the method by comparing the positive indications
for DM scattering from CDMS-Si and the annual modulation in DAMA with
the limits from the LUX and SuperCDMS experiments. For DAMA we confirm
previous results \cite{HerreroGarcia:2012fu} and we obtain very low 
joint probabilities of $p_{\rm joint} \lesssim 10^{-7}$ for SuperCDMS and 
$p_{\rm joint} \simeq 10^{-8}$ for LUX. 

In the case of CDMS-Si we face the situation that the signal itself is
rather weak, consisting of only 3 events with an expected background
of 0.62 events. When only the information of the number of events is
used the CL for a signal being present is very low (around 97.4\%),
and hence a possible exclusion of the signal from other experiments is
possible at best at the same very modest CL. For this reason we have
identified a method to take into account energy information. An
important observation is that the three events cluster at relatively
low energies, as expected from a DM signal, compared to the more
broadly distributed background. Therefore, we consider the so-called
signal length, defined as the expected number of events in the energy
interval between the events with the lowest and highest energies. We
can calculate the probability of obtaining a signal length equal or
less than the one obtained in CDMS-Si and use it to get the joint
probability with constraints from LUX and SuperCDMS. We find CDMS-Si
and SuperCDMS being consistent with a probability of 4\% for $m_\chi
\simeq 5$~GeV, decreasing to 0.5\% for $m_\chi \simeq 40$~GeV. For
$m_\chi \gtrsim 14$~GeV, LUX provides a strong bound leading to a
joint probability basically given by the probability of the background
only hypothesis of 0.17\%.

We have also applied our method to the case of isospin violating
interactions, where for a specific choice of the DM coupling strengths
to neutrons and protons the scattering cross section of the
experiments providing upper limits can be significantly suppressed
relative to the one relevant for the experiment reporting a
signal. For Xe (Ge) the maximum suppression occurs for $f_n/f_p =
-0.7$ ($-0.8$). A careful choice of $f_n/f_p$ allows for better
compatibility of CDMS-Si with LUX and SuperCDMS (at relatively small
DM masses), whereas for DAMA for any $f_n/f_p$ the joint probability with 
at least one of the other experiments remains small.

The signal length method used in this work to take into account energy
information is one particular observable which turns out to be useful
to test the CDMS-Si signal. It is inspired by the popular maximum-gap
method~\cite{Yellin:2002xd}, and in addition to the total number
  of events it takes into account a second observable (the ``signal
  length'') given by the properly defined distance between the two
  events with the lowest and highest energy. It has the advantage
that the relevant probability can be analytically calculated and is
relatively simple, see Eq.~\eqref{eq:prob}.  The signal length method
provides a goodness-of-fit test returning a probability for the actual
experimental outcome to occur under a given hypothesis. It is useful
in the case of signals consisting of few (but at least two) events.
Further discussion of the method can be found in
App.~\ref{app:SL}. While the signal length turns out to be powerful in
the case of CDMS-Si we do not exclude the possibility that in other
situations different observables might also be identified and used in
a similar fashion as the signal length.

Let us stress that the probabilities obtained by our method are
actually upper bounds on the joint probability. In several steps in
our calculations we use inequalities or maximization, for instance in
Eq.~\eqref{eq:muvm} to set an upper bound on the halo integral, in
calculating the relevant probability for the signal length in
section~\ref{sec:method-signal-length}, or in maximizing the joint
probability $p_A p_B$ in Eqs.~\eqref{eq:pjoint1} or
\eqref{eq:pjoint2}. The true probability of obtaining the two
experimental results will actually be lower than the value returned by
our method.

Let us briefly compare our results to the ones obtained by Feldstein
and Kahlhoefer in Ref.~\cite{Feldstein:2014ufa}. They consider the
joint likelihood function for CDMS-Si, SuperCDMS, and LUX, optimized
with respect to all possible DM halo configurations and the DM mass.
In addition a constraint on the galactic escape velocity is imposed in
their likelihood. They obtain a best fit DM mass of 5.7~GeV and
determine the p-value of the fit by Monte Carlo method as 0.44\%.
From our Fig.~\ref{fig:prob-Si} we find at $m_\chi = 5.7$~GeV a joint
probability of CDMS-Si and SuperCDMS of 3.6\%, whereas the LUX
constraint is absent at those low DM masses. We stress that the
approaches are quite different, leading to different statistical
statements. Hence a direct quantitative comparison is difficult.  The
likelihood method uses a maximum of information from each event, so
even the single event tested by LUX for $m_\chi = 5.7$~GeV provides
some constraint. Our method uses energy information in a more
condensed way (via the signal length) and is more conservative in
several respects. It has the advantage of providing directly a
probability for how likely the experimental outcome is under the DM
hypothesis, without the need of Monte Carlo simulations.

\subsection*{Acknowledgements}

We thank Brian Feldstein and especially Felix Kahlhoefer for comments
on the manuscript and very useful discussions. We acknowledge support
from the European Union FP7 ITN INVISIBLES (Marie Curie Actions,
PITN-GA-2011-289442). N.B.\ thanks the Oskar Klein Centre and the CoPS
group at the University of Stockholm for hospitality during her
long-term visit.

\appendix
\section{The signal length method}
\label{app:SL}

In this appendix we provide some details on the signal length 
method.  In App.~\ref{app:prob} we derive the joint probability
$P_{\rm SL}(N^{{\rm obs}}, \Delta|\mu)$ from Eq.~\eqref{eq:prob} of
obtaining $N^{{\rm obs}}$ or more events and a signal length of size
$\Delta$ or smaller. We discuss some properties of the SL test in
App.~\ref{app:CDMS}, where we also demonstrate that the SL method can
be used to obtain allowed regions in DM mass and scattering cross
section if a specific halo is adopted for the example of the CDMS-Si
data. A general discussion of the method follows in
Sec.~\ref{app:discussion}.

\subsection{Probability derivation}
\label{app:prob}

Let us denote by $dN/dE_{nr}$ the expected event spectrum for a given DM
model, DM halo, and background. The expected number of events between two
energies $E_1$ and $E_2$ is then given by
\be
N_{[E_1,E_2]} = \int_{E_1}^{E_2} dE_{nr} \, \frac{dN}{dE_{nr}} \,.
\ee
The value of $N_{[E_1,E_2]}$ is invariant under a change of
variable. In particular, we can always use a new variable $x$ with $dx
= dN$, such that the distribution of $x$ is constant and equal to
unity in the interval $[0, \mu]$, where $\mu$ is the expected number of
events in the total energy interval \cite{Yellin:2002xd}. The expected
number of events in a given energy interval is simply $N_{[E_1,E_2]} =
x_2 - x_1$.

Hence the problem reduces to the following. Assume $n$ independent
random numbers $x_n$, uniformly distributed in the interval $[0,
  \mu]$.  We order the events as $x_1 < x_2 < ... < x_n$. The ``signal
length'' is then given by $\Delta = x_n - x_1$. We want to calculate
the probability of obtaining a signal length less than $\Delta$ for given
$n$, $P({\rm SL} < \Delta|n) = 1 - P({\rm SL} \ge \Delta|n)$.
We calculate $P({\rm SL} \ge \Delta|n)$ as (probability of $x_1 \in [0, \mu-\Delta]$) times
(probability of $x_n \in [x_1+\Delta,\mu]$) times (probability of all $x_2,...,x_{n-1}$
between $x_1$ and $x_n$) times a combinatorial factor:
\be
P({\rm SL} \ge \Delta|n) = 2 
\left(\begin{array}{c} n \\ 2 \end{array}\right) \frac{1}{\mu^n}
\int_0^{\mu-\Delta} dx_1 \int_{x_1 + \Delta}^\mu dx_n \, (x_n - x_1)^{n-2} \,.
\ee
The binomial coefficient 
\be
\left(\begin{array}{c} n \\ 2 \end{array}\right) = \frac{n!}{(n-2)! 2!} = \frac{n(n-1)}{2}
\ee
corresponds to the number of possibilities to pick $2$ events out of
$n$ (the ones we call $x_1$ and $x_n$), and the factor 2 is needed
because of the two possibilities of either of them being the smaller
or the larger. The factor $1/\mu^n$ ensures that the probability of
finding each event between 0 and $\mu$ is normalized to 1. The integrals are easily calculated
and we find
\be
P({\rm SL} < \Delta|n) = 1 - P({\rm SL} \ge \Delta|n) = 
n \left(\frac{\Delta}{\mu}\right)^{n-1} - (n-1)\left(\frac{\Delta}{\mu}\right)^n \,.
\label{eq:probx}
\end{equation}
The corresponding probability distribution function (pdf) for the signal length $\Delta$ 
is obtained by differentiating Eq.~\eqref{eq:probx} as
\begin{equation}
f( \Delta |n) = n(n-1) \frac{1}{\mu} \left( \frac{\Delta}{\mu} \right)^{n-2} \left(1- \frac{\Delta}{\mu} \right) \,,
\end{equation}
which is defined for $n \ge 2$ and $\Delta \le \mu$.

The joint pdf for the number of events $n$ and the signal length $\Delta$ for a given $\mu$ is 
obtained from Bayes' theorem as
\begin{equation}\label{eq:pdf}
  f(n, \Delta) = f(\Delta | n) f(n) = 
 \frac{1}{(n-2)!} \, e^{-\mu} \Delta^{n-2} (\mu - \Delta) \,.
\end{equation}
We have used that $n$ is Poisson distributed with mean $\mu$, and 
$f(n, \Delta)$ is defined for $n \ge 2$ and $\Delta \le \mu$. 
The normalization is such that 
\begin{equation}
  \sum_{n=2}^\infty \int_0^\mu d\Delta f(n, \Delta) = \mathcal{P}_\mu(n\ge 2) \,, 
\end{equation}
where $\mathcal{P}_\mu(n\ge 2) = 1 - e^{-\mu}(1+\mu)$ is the 
Poisson probability for obtaining $n\ge 2$ for Poisson mean $\mu$. Hence the normalized pdf
is given by
\begin{equation}\label{eq:pdf-norm}
  \tilde f(n,\Delta) \equiv \frac{f(n,\Delta)}{\mathcal{P}_\mu(n\ge 2)} \,.
\end{equation}
It describes the distribution of $n$ and $\Delta$ given that $n\ge 2$ events have been observed.

Suppose an experiment observes $N^{\rm obs}$ events. Then the
joint probability to obtain a number of events equal or larger than
$N^{\rm obs}$ and a signal length equal or smaller than $\Delta$ for a given $\mu$ is
\begin{align}
  P_{\rm SL}(N^{\rm obs}, \Delta|\mu) &= \sum_{n=N^{{\rm obs}}}^\infty \int_0^\Delta d\Delta' f(n, \Delta') \\
               &= e^{-\mu} \sum_{n=N^{{\rm obs}}}^\infty \frac{1}{n!} [n\mu \Delta^{n-1} - (n-1)\Delta^n] \,.
\label{eq:prob-app}
\end{align}
Eq.~\eqref{eq:prob-app} corresponds to the expression in
Eq.~\eqref{eq:prob} used in our analysis of DM direct detection data.

\subsection{Properties of the SL and application to CDMS-Si data}
\label{app:CDMS}

\begin{figure}
\begin{center}
 \includegraphics[width=0.55\textwidth]{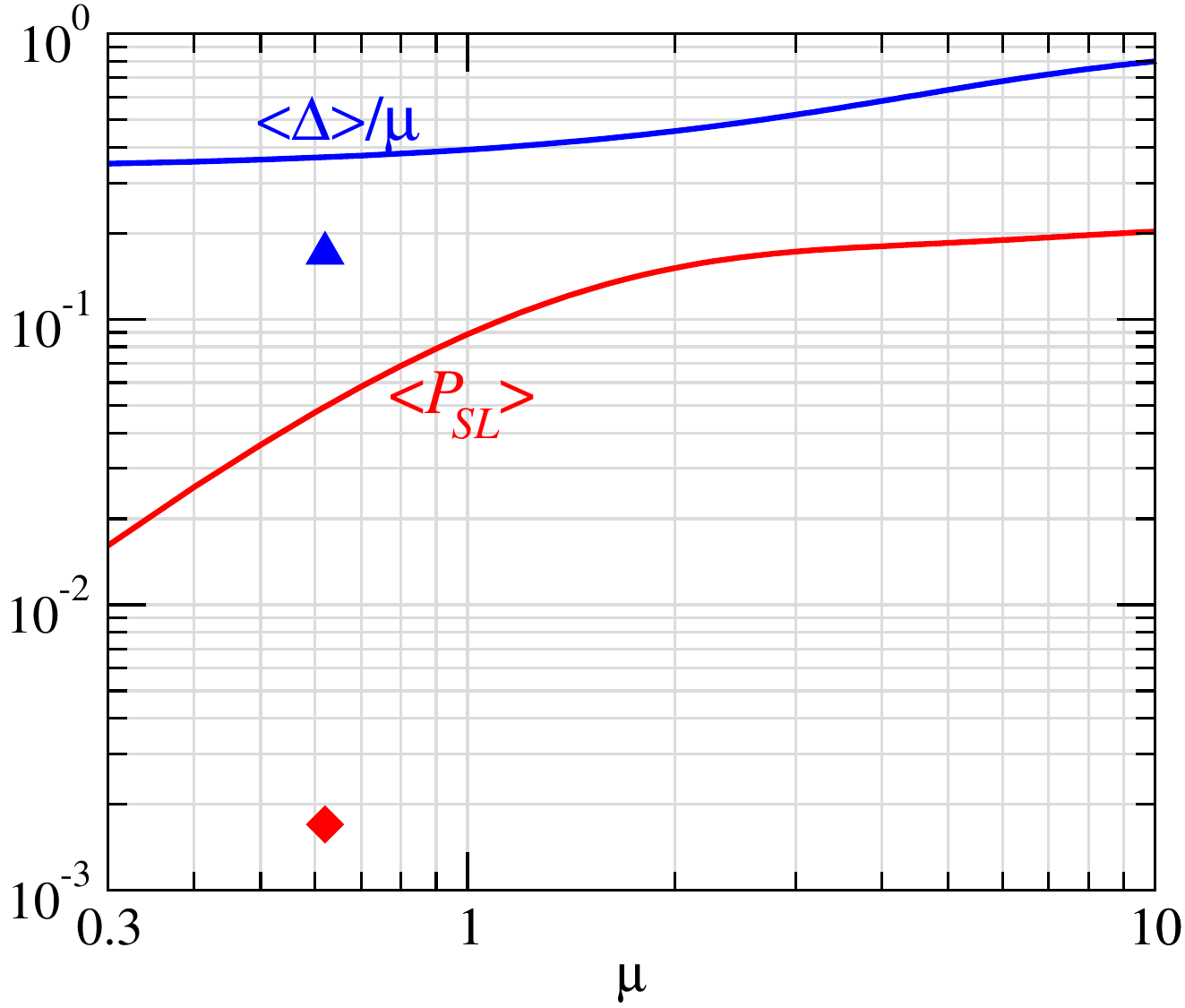}
\caption{\label{fig:SL-mean} Expectation values of the signal length
  $\langle\Delta\rangle / \mu$ and of the signal length probability
  $\langle P_{\rm SL}\rangle$ as a function of $\mu$. The blue triangle
  and the red diamond indicate the values of $\Delta/\mu$ and
  $P_{\rm SL}$, respectively, obtained for the background-only hypothesis
  for CDMS-Si for which $\Delta = 0.104$ and $\mu = 0.62$.}
\end{center}
\end{figure}

Using the normalized pdf, Eq.~\eqref{eq:pdf-norm}, we can calculate the expectation value of any 
quantity $X$ which depends on the random variables $n$ and $\Delta$:
\begin{equation}\label{eq:expectation}
  \langle X \rangle = \sum_{n=2}^\infty \int_0^\mu d\Delta \, X \, \tilde f(n, \Delta) \,.
\end{equation}
For instance we can calculate the expected value of the signal
length $\Delta$. The blue curve in Fig.~\ref{fig:SL-mean} shows
$\langle\Delta\rangle / \mu$ as a function of the expected number of
events $\mu$. The probability $P_{\rm SL}$ will be small if a SL much
smaller than $\langle\Delta\rangle$ is observed. The blue triangle
shows the value of $\Delta/\mu$ corresponding to the background-only
hypothesis of CDMS-Si, with $\Delta = 0.104$ and $\mu = 0.62$.

We can use Eq.~\eqref{eq:expectation} to calculate also the expected
value of $P_{\rm SL}$ itself, considered as a function of $\Delta$ and
$n$:
\begin{equation}
  \langle P_{\rm SL} \rangle = \sum_{n=2}^\infty \int_0^\mu d\Delta \, P_{\rm SL}(n, \Delta|\mu) \, 
   \tilde f(n, \Delta) \,.
\end{equation}
This is shown by the red curve in Fig.~\ref{fig:SL-mean}. We observe
that for $\mu \lesssim 2$ we expect that $P_{\rm SL}$ becomes small,
because it is unlikely to obtain at least 2 events. For $\mu \gtrsim
2$ we find $\langle P_{\rm SL} \rangle \simeq 0.2$. Hence, whenever the
observed value for $P_{\rm SL}$ is much smaller than 0.2 the experimental
outcome can be considered to be unlikely. The red diamond in
Fig.~\ref{fig:SL-mean} indicates the value $P_{\rm SL} = 0.17\%$ obtained
for the background-only hypothesis for CDMS-Si, which is significantly
smaller than the expectation.

\begin{figure}
\begin{center}
 \includegraphics[width=0.55\textwidth]{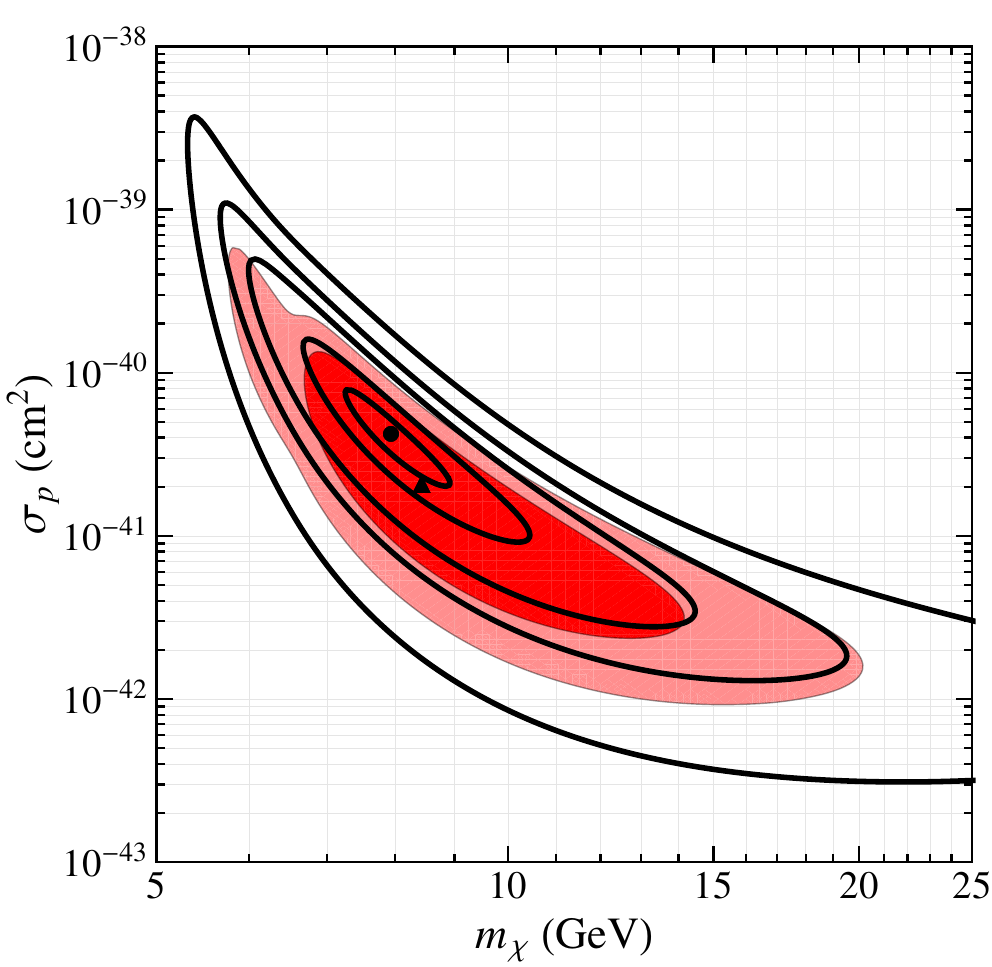}
\caption{\label{fig:SHM} Regions in the plane of DM mass and
  scattering cross section from CDMS-Si data assuming the so-called
  standard halo model. The black curves show equal probability
  contours using the signal length method corresponding to $P_{\rm SL} =
  0.01, 0.05, 0.1, 0.2, 0.25$, from outer to inner contours. The black
  dot indicates the point of highest probability. The shaded regions
  (black triangle) correspond to 68\% and 90\%~CL regions (best fit
  point) using an extended maximum likelihood analysis.}
\end{center}
\end{figure}

If on top of the backgrounds a specific DM hypothesis is specified, the
SL method provides a way to quantify how likely the experimental
outcome is under this hypothesis. In Fig.~\ref{fig:SHM} we show the
results of such an analysis for CDMS-Si data, assuming spin-independent elastic
DM--nucleon interactions and the so-called standard halo model for the
local DM velocity distribution.\footnote{We adopt the conventional
  Maxwellian velocity distribution with $\bar v = 220$~km/s, truncated
  at the escape velocity of $v_\mathrm{esc} = 544$~km/s. The velocity
  of the Sun in galactic coordinates is (10,233,7)~km/s and we assume
  a local DM density $\rho_\chi = 0.3$~GeV/cm$^3$.} Then, for each point
in the plane of DM mass and scattering cross section we can calculate
$\mu$, the SL $\Delta$, and $P_{\rm SL}$. The black curves in the figure
show contours of constant values of $P_{\rm SL}$ between 0.01 and
0.25. The black dot indicates the point of highest probability, which
is $P_{\rm SL} = 0.266$, i.e., close to the expectation value (compare to
Fig.~\ref{fig:SL-mean}).

The contours of constant $P_{\rm SL}$ can be compared to more traditional
regions obtained from an extended maximum likelihood analysis of
CDMS-Si data, shown by the shaded regions in Fig.~\ref{fig:SHM}.  This
analysis is similar to the ones performed in \cite{Bozorgnia:2013pua,
  Frandsen:2013cna} and leads to allowed regions in good agreement
with the ones obtained by the CDMS-Si collaboration
\cite{Agnese:2013rvf}. We observe very good agreement of the regions
between the two methods, although we stress the different
meanings. The SL method provides regions where the experimental
outcome is likely (in the sense of a goodness-of-fit test), while
the maximum likelihood method leads to confidence regions in the
parameter space under the hypothesis that the DM interpretation is
correct (i.e., regions relative to the best fit point).

\subsection{Discussion}
\label{app:discussion}

The SL method can be used to quantify how likely a given experimental
outcome is for a specified hypothesis. It works well for low event
numbers (but at least 2 events are needed). It is based on the predicted number of events
between the energy of the lowest and highest event (the ``signal
length'' $\Delta$) as well as on the predicted number of events in the
full energy range $\mu$. In some sense the SL method is based on
  two ``bins'', the total energy interval and the one between the
  lowest and highest event. The size of the $\Delta$ bin is determined
  by the data and it is actually the bin size which contains the
  relevant information used to calculate the probability. The full
probability distribution of the expected events has to be known to
calculate $\Delta$ and $\mu$, similar to the case of a likelihood
analysis. While the SL method provides a goodness-of-fit test,
returning an absolute probability, a maximum likelihood analysis is
based on the relative likelihood of parameter points with respect to
the best fit point (assuming that the model itself is correct).

The SL method is powerful in the case of few events to evaluate signal
versus background-only hypotheses which have a distinct energy shape,
for instance a peaked signal versus a broader distribution of
background. The method becomes not very useful in case of many
events. In this case the two numbers $\Delta$ and $\mu$ provide only
limited information on the detailed event spectrum and in such a case
alternative methods will be more powerful, for example binned $\chi^2$
goodness-of-fit, un-binned Kolmogorov-Smirnov shape test, or
likelihood ratio tests.

The relevant probability $P_{\rm SL}$ for the SL test has a relatively
simple expression, see Eq.~\eqref{eq:prob-app}, which can easily be
evaluated numerically. Thanks to this simple form and the fact that
$P_{\rm SL}$ depends only on $\Delta$ and $\mu$, the SL method is also
useful if the hypothesis is constrained by other data or consistency
requirements. Maximization with respect to $\Delta$ and $\mu$ provides
then an upper bound on the probability, as demonstrated explicitly in
Sec.~\ref{sec:method-signal-length}. Similar methods may be used also
to take into account uncertainties in the predicted spectrum, for
instance uncertainty on the expected background.

\bibliographystyle{my-h-physrev.bst}
\bibliography{./refs}

\end{document}